\documentclass[epj]{svjour}
\usepackage{graphics}

\usepackage{amsmath, amssymb}
\usepackage{multirow}
\usepackage{longtable}
\usepackage{color}
\usepackage{hyperref}

\usepackage{slashed}

\DeclareMathOperator*{\sumint}{%
\mathchoice%
  {\ooalign{$\displaystyle\sum$\cr\hidewidth$\displaystyle\int$\hidewidth\cr}}
  {\ooalign{\raisebox{.14\height}{\scalebox{.7}{$\textstyle\sum$}}\cr\hidewidth$\textstyle\int$\hidewidth\cr}}
  {\ooalign{\raisebox{.2\height}{\scalebox{.6}{$\scriptstyle\sum$}}\cr$\scriptstyle\int$\cr}}
  {\ooalign{\raisebox{.2\height}{\scalebox{.6}{$\scriptstyle\sum$}}\cr$\scriptstyle\int$\cr}}
}

\begin{document}
\title{ Driving chiral phase transition with ring diagram }
\author{Pok Man Lo\inst{1}, 
Micha\l\, Szyma\'nski\inst{1},
Krzysztof Redlich\inst{1},
Chihiro Sasaki\inst{1}
}                     
\offprints{}          
\institute{Institute of Theoretical Physics, University of Wroclaw,
PL-50204 Wroc\l aw, Poland}
\date{Received: date / Revised version: date}
%
\abstract{
    We study the dressing of four-quark interaction by the ring diagram, and its feeding back to the quark gap equation, in an effective chiral quark model. Implementing such an in-medium coupling naturally reduces the chiral transition temperature in a class of chiral models, and is capable of generating the inverse magnetic catalysis at finite temperatures.  We also demonstrate the important role of confining forces, via the Polyakov loop, in a positive feedback mechanism which reinforces the inverse magnetic catalysis.
} 
\titlerunning{Driving chiral phase transition with ring}
\maketitle
\section{Introduction}


A robust description of chiral symmetry restoration and its manifestation in
a medium of partially deconfined quarks and gluons is essential to making
progress in understanding the properties of QCD matter under extreme conditions, such as those created in the laboratory during the ultra-relativistic heavy-ion collisions or fill the core of neutron stars. 

Effective models are a flexible exploratory tool to study a dynamical system.
One of the advantage is the ability to temporarily include (or suppress) a certain class of interactions or diagrams and examining the effect in isolation. 
One can also gain insights on the values of phenomenological parameters used and examine their connections to the properties of underlying constituents. 
These make this approach useful to complement the more powerful numerical methods such as lattice QCD (LQCD). 

In this paper we study the in-medium dressing of the four-quark interaction by
resumming a class of ring diagrams within an effective chiral quark model.  
Screening of the potential by ring diagram finds the most famous application in regulating 
the long-range Coulomb forces in an electron gas~\cite{GellmannBrueckner,ElectronGas}. 
(See Refs~\cite{Mattuck,Leeuwen_2013} for discussions in condensed matter theory.)
Here we shall see that it reveals rich features of the QCD phase diagram in an
effective model~\cite{conf,polarNJL,Ayala01,Ayala:2021nhx}.

In a previous work~\cite{new} we have demonstrated that polarization provides a natural mechanism to connect the large transition temperature scale ($T_d \approx 270$ MeV) in a pure gauge theory and that of chiral symmetry restoration
($T_{pc} \approx 156.5$ MeV) in the presence of light, dynamical quarks.
It also drives the phenomenon of inverse magnetic
catalysis~\cite{Andersen:2014xxa,Miransky:2015ava} at finite temperatures, 
i.e. the chiral condensate decreases more rapidly with temperature in the presence of a magnetic field.

In this work we explore further theoretical issues of the proposed model.
We shall revisit the chiral condensate and in addition examine how the Polyakov
loops are influenced by the ring diagram.
We shall also elucidate some details in the calculation of the polarization tensors in the vector,
scalar and pseudoscalar channels. 
While the calculation of ring diagram with a given quark mass is well
known~\cite{Klevansky1992,Jankowski:2009kr,Avancini:2015ady,Zhang:2016qrl}, 
its feeding back to the quark gap equation for consistent solution, 
and thereby including the backreaction, is usually not performed in studies of effective quark model.
As we shall see, it is precisely this extra step that leads to a substantial change in the temperature and
magnetic field dependences of quark condensate, and is demonstrated to improve the description of
many aspects of QCD phase diagram. A merit of the current scheme is that there
is no need to introduce an artificial tuning of $T_d$ parameter in the gluon sector,
nor the need to introducing an explicit $B$-dependent coupling, as advocated in
Refs.~\cite{mediumG1,mediumG2}. Instead, the mechanism serve as a tentative
explanation for such a medium dependence.

\section{Chiral quark model with dressed interaction}

We begin with a brief review of the theoretical model: 
an effective chiral quark model motivated from the Coulomb Gauge QCD~\cite{conf,Govaerts:1983ft,Kocic:1985uq,Hirata:1989qp,Alkofer:1989vr,Schmidt:1995gea,Reinhardt:2017pyr,Quandt:2018bbu}. 
The Lagrangian density reads:
\begin{equation}
    \label{eq:model}
    \begin{split}
    \mathcal{L}(x) &= \bar{\psi}(x) \, (i \slashed{\partial}_x - m) \, \psi(x) \\
    & \quad - \frac{1}{2} \,
    \int d^4 y \,  \rho^a(x) \, V^{ab}(x, y) \, \rho^b(y)
    \end{split}
\end{equation}
where $\rho^a(x) = \bar{\psi}(x) \gamma^0 T^a \psi(x)$ is the color quark current and
$T^a$ is a generator of the $SU(N_c)$ symmetry group, with $a = 1, 2, \ldots, N_c^2-1$. 
For the class of model where the interaction potential $V$ is
instantaneous and color-diagonal, i.e.,

\begin{equation}
    \label{eq:instpot}
    V^{ab}(x, y) \rightarrow \delta^{ab} \times \delta(x^0-y^0) \,
    V(\vec{x}-\vec{y}),
\end{equation}
the gap equation for the dynamical quarks has been 
derived~\cite{conf}. 
The leading order result can be summarized as follows:

\begin{equation}
        S^{-1}(p) = \slashed{p} - m - \Sigma(p)
        \label{eq:sinv}
\end{equation}
where
\begin{equation}
    \Sigma(p) = C_F \, \int \frac{d^4 q}{(2 \pi)^4} \, V(\vec{p}-\vec{q}) \, i \, \gamma^0 S(q) \gamma^0.
        \label{eq:sfquark}
\end{equation}
The constant $C_F = \frac{N_c^2-1}{2 N_c}$ is introduced via 
the quadratic Casimir operator

\begin{equation}
    \sum_{a=1}^{N_c^2-1} \, T^a T^a = C_F \, \mathcal{I}_{N_c \times N_c}.
\end{equation}
The solution to the gap equation~\eqref{eq:sinv} can be parametrized as

\begin{equation}
        S^{-1}(p) = A_0(p) \, p^0 \gamma^0 - A(p) \, \vec{p} \cdot \vec{\gamma} - B(p)
\end{equation}
with the quark dressing functions, to be determined self-consistently, 
given by

\begin{equation}
    \begin{split}
        A_0(p) &= 1 \\
        A(p) &= 1 + C_F \, \int \frac{d^3 q}{(2 \pi)^3} \, V(\vec{p}-\vec{q}) \,
        \frac{A(q) \, \hat{p} \cdot \hat{q}}{2 \tilde{E}(q)} \, \Theta \\
        B(p) &= m + C_F \, \int \frac{d^3 q}{(2 \pi)^3} \, V(\vec{p}-\vec{q}) \,
        \frac{B(q)}{2 \tilde{E}(q)} \, \Theta \\
        \Theta &= 1 - 2 \, N_{\rm th}(\tilde{E}),
    \end{split}
    \label{eq:gap}
\end{equation}
where $\tilde{E}(q) =  \sqrt{ A(q)^2 \vec{q}^2 + B(q)^2} $ is a generalized
energy function of the dynamical quarks, 
$N_{\rm th}(E) = \frac{1}{e^{\beta E}+1} $ is the Fermi-Dirac distribution. 

Considering an instantaneous gluon potential~\eqref{eq:instpot} 
means that there is no $p_0$ dependence in the gap equations. 
The remaining dependence on the 3-momentum $\vec{p}$ disappears when considering a contact
interaction: taking
\begin{equation}
V(\vec{p}-\vec{q}) \rightarrow V_0
\end{equation}
in Eq.~\eqref{eq:gap} immediately forces $A(p) = 1$, and the quark mass function reduces to

\begin{equation}
    \label{eq:cgap}
        M = m + C_F \, V_0 \, \int \frac{d^3 q}{(2 \pi)^3} \,
        \frac{M}{2 \sqrt{q^2 + M^2}} \Theta.
\end{equation}
This is of the same form as the familiar result for quark mass ($N_f$ flavors) in the
model~\cite{Klevansky1992} of Nambu and Jona-Lasinio (NJL), with the identification 

\begin{equation}
      \label{eq:match}
      C_F \, V_0 \leftrightarrow 4 \, N_c \, N_f \, (2 \, G_{\rm NJL}).
\end{equation}
In fact the present model provides a more natural starting point as an effective
model of QCD:
First, it closely mimics the quark-gluon interactions of QCD
by implementing a vector nature of the four-quark interactions 
originated from a gluon exchange, 
both in the color and the Dirac space.
Note that an effective interaction in the scalar-scalar channel is also generated from such
vector-vector (from Fock-type exchange), giving rise to a spontaneous chiral symmetry breaking.
This may also be understood from a Fierz transformation~\cite{Buballa:2003qv} of the original Lagrangian in
Eq.~\eqref{eq:model}: a vector-vector interaction can generate scalar-scalar type
interactions (and vice versa). 
Second, it makes possible a systematic improvement on the quark potential 
by taking into account features of gluon propagators, e.g. momentum dependence.

The generalization of the model~\eqref{eq:cgap} 
to include an in-medium dressing of the interaction potential $V_0$ 
via the polarization tensor $\Pi_{00}$~\cite{conf} proceeds by:

\begin{equation}
    \tilde{V_0}^{-1} = {V_0}^{-1} - \frac{1}{2} \, N_f \, \Pi_{00}
    \label{eq:dress}
\end{equation}
where
\begin{equation}
    \label{eq:pi00}
    \begin{split}
        \Pi_{00}(p^0, \vec{p}) &= \frac{1}{\beta} \, \sumint {\rm Tr} \left( \gamma^0
        S(q) \gamma^0 S(q+p) \right).
    \end{split}
\end{equation}
Here $\sumint$ denotes a Matsubara sum over the fermionic frequencies
($\omega_n = (2n+1) \, \pi/\beta$), and an integral over the momenta $d^3 q$. 
In this work we work only in the static, vanishing momentum limit of the ring and thus
$p^0 =0, \vec{p} \rightarrow \vec{0}$ are eventually taken in the calculation.

Eq.~\eqref{eq:dress} can be understood as the dressing of the gluon propagator 
by the Debye mass. The factor of $\frac{1}{2}$ in Eq.~\eqref{eq:dress} originates 
from the color structure, i.e. ${\rm Tr} \, T^a T^b = \frac{1}{2} \, \delta^{ab}$, 
and is essential to reproduce the known result~\cite{Kapusta:2006pm}
of the perturbative Debye mass for QCD, instead of QED.

We choose to work in an effective model with quarks and gluons as the degrees of
freedom. According to quark-hadron duality one should be able to include the hadron effects by including, 
and iterating, multi-particle interactions among quarks and gluons. 
The appearance of the higher order terms in a quark-based picture, however, is
different from those constructed out of mesons. There is no one-to-one mapping without
further approximation~\cite{polarNJL,Ayala01,Skokov:2011ib,Andersen:2013swa}. 
Even in the usual NJL model~\cite{Klevansky1992}, where a Hubbard–Stratonovich
transformation is used to introduce the meson fields, the kinetic
terms~\cite{Hamazaki:1994rf} of mesons are not formally derived. 
In this work we shall explore the effect of quark loops and their feeding
back to the quark gap equation, thus going beyond the standard mean-field
treatment.

In many studies, polarization tensors are computed with the fermion propagator determined 
from a leading order mean-field gap equation such as Eq.~\eqref{eq:cgap}. 
The use of $\tilde{V_0}$, in lieu of $V_0$, amounts to implementing a back-reaction of the
fermion loops to the fermionic gap equation. In the language of condensed matter
theory, the scheme is similar to an iteration of GW-scheme~\cite{Mattuck,Leeuwen_2013} 
with polarization insertions but without vertex corrections. 
This effectively dresses the four-quark interaction and can substantially modify aspects of
chiral phase transition, such as driving the phenomenon of inverse magnetic catalysis.

\section{Polarization tensors}

Many observables within an NJL-like model can be understood in terms of the
following integrals~\cite{Klevansky1992,Zhang:2016qrl}: 

\begin{equation}
    \label{eq:I_collect}
    \begin{split}
        I_0 &= \frac{1}{\beta} \, \sumint \,
        \frac{1}{\omega_n^2 + E_1^2}, \\
        I_1(p^0, \vec{p}) &= \frac{1}{\beta} \, \sumint \,
        \frac{1}{\omega_n^2 + E_1^2} \, \frac{1}{(\omega_n - i \, p^0)^2 +
        E_2^2}, \\
        I_2(p^0, \vec{p}) &= \frac{1}{\beta} \, \sumint \,
        \frac{1}{\omega_n^2 + E_1^2} \,
        \frac{\vec{q}^2 + \vec{q} \cdot \vec{p}}{(\omega_n - i \, p^0)^2 + E_2^2}.
    \end{split}
\end{equation}
Note that the (constituent) quark mass dependence enters via 
$E_i = \sqrt{\vec{q}_i^2 + M^2}$, where $\vec{q}_1 = \vec{q}$ and 
$\vec{q}_2 = \vec{q}+\vec{p}$.
These integrals can be decomposed into a UV-divergent vacuum piece and a
finite temperature piece. For example, $I_0$ can be written as

\begin{equation}
    \label{eq:decompo}
        I_0 = I_0^{\rm vac} + I_0^T.
\end{equation}
The first piece requires regularization, e.g., by a 3D regulator $\mathcal{R}_{\rm 3D}(q) =
e^{-q^2/\Lambda^2}$:

\begin{equation}
    I_0^{\rm vac}  \rightarrow
    \int \frac{d^3 q}{(2 \pi)^3} \, \frac{1}{2 E_1} \, \mathcal{R}_{\rm 3D}(q).
\end{equation}
Alternatively, one can choose a 4D cutoff scheme:

\begin{equation}
    I_0^{\rm vac}  \rightarrow
    \int \frac{d^4 q_E}{(2 \pi)^4} \, \frac{1}{q_E^2 + E_1^2} \, \mathcal{R}_{4D}(q),
\end{equation}
or a Schwinger proper-time regularization scheme:

\begin{equation}
    I_0^{\rm vac}  \rightarrow
    \int_{1/\Lambda^2}^{\infty} \, \frac{dt}{(16 \pi^2)} \, \frac{1}{t^2} \, e^{-M^2 \, t}.
\end{equation}
The finite temperature piece, on the other hand, requires no regularization, and is given by

\begin{equation}
        I_0^T = \int \frac{d^3 q}{(2 \pi)^3} \, \frac{-1}{2 E_1}
        \times 2 N_{\rm
    th}(E_1),
\end{equation}
Under a general regularization scheme, the finite temperature piece of $I_0$ can be defined
in a regularization independent manner~\cite{regulators,Lo:2013lca} by 

\begin{equation}
    \label{eq:ftextract}
    I_0^T = \lim_{\Lambda \rightarrow \infty} \,
    \left(I_0(T, \Lambda) - I_0(T \rightarrow 0, \Lambda)
    \right).
\end{equation}
Similar analysis can be applied to $I_1$ and $I_2$. The results are:

\begin{equation}
    \begin{split}
        I_1(p^0, \vec{p}) &= \int \frac{d^3 q}{(2 \pi)^3} \, \frac{-1}{4 E_1 E_2} \times (Q_1 + Q_2), \\
        Q_1 &= (\mathcal{R}(q)-N_1-N_2) \, \times \\
        &\quad (\frac{1}{p^0-E_1-E_2}-\frac{1}{p^0+E_1+E_2} ) \\
        Q_2 &= (N_1-N_2) \, \times \\
        &\quad (\frac{1}{p^0-E_1+E_2}-\frac{1}{p^0+E_1-E_2})
    \end{split}
    \label{eq:I1}
\end{equation}

and
\begin{equation}
    \begin{split}
        I_2(p^0, \vec{p}) &= -\int \frac{d^3 q}{(2 \pi)^3} \, \frac{\vec{q}^2 + \vec{q} \cdot \vec{p}}{4 E_1 E_2} \times  \, (Q_1 + Q_2), \\
    \end{split}
    \label{eq:I2}
\end{equation}
where $N_i = N_{\rm th}(E_i)$.

The merit of studying these expressions~\eqref{eq:I_collect} 
is that various results of the model can be written in terms of them. 
For example, the gap equation in Eq.~\eqref{eq:cgap} can be neatly expressed as

\begin{equation}
        M = m + C_F \, V_0 \, M \times I_0.
    \label{eq:cgap2}
\end{equation}
The chiral condensate (per flavor) is given by 

\begin{equation}
        \langle \bar{\psi} \, \psi \rangle = - 4 N_c M \times I_0.
\end{equation}
Moreover, the pion decay constant ($N_f=2$) can be estimated from the low energy limit of $I_1$ by
\begin{equation}
    f_\pi^2 \approx 4 N_c M^2 \times I_1(p^0 \rightarrow 0, \vec{p}=\vec{0}).
    \label{eq:fpi}
\end{equation}

In this work the four-quark coupling $\tilde{V}_0$ in Eq.~\eqref{eq:dress} is dressed 
by the $\Pi_{00}$ polarization tensor evaluated at the static limit.
Note that the full polarization tensor $\Pi_{00}(p^0, \vec{p})$ can also be expressed 
in terms of integrals in Eq.~\eqref{eq:I_collect} as

\begin{equation}
        \Pi_{00}(p^0, \vec{p}) = 4 \left( -I_0 + (-\frac{1}{2} p^2 + 2 M^2) \, I_1 + 2 I_2 \right).
        \label{eq:pi00_2}
\end{equation}
In the zero-temperature and static limit the expression in Eq.~\eqref{eq:pi00_2} vanishes. 
This is a familiar result in the Hard-Thermal-Loop (HTL) study~\cite{lebellac}, where a further $M \rightarrow 0$ limit is
implicitly taken. The relation remains true for a general $M$, as one can directly verify

\begin{equation}
    \Pi_{00}(0, \vec{p}=\vec{0}) \propto \left( -I_0 + 2 M^2 \, I_1(0, \vec{0}) + 2 I_2(0,
    \vec{0}) \right),
\end{equation}
and the second and third terms add up to $I_0$, exactly canceling the first term. 
See Eqs.~\eqref{eq:I1} and~\eqref{eq:I2}.

For the finite temperature part, 
besides a direct numerical evaluation of Eq.~\eqref{eq:pi00_2}, 
an alternative convenient method to obtain the result~\cite{new} is through 
a formal relation to the thermal pressure of a free (single species) fermion gas at finite
temperature and vanishing chemical potential:

\begin{equation}
    \begin{split}
        \Pi_{00}(p^0=0, \vec{p}\rightarrow \vec{0}) &= \frac{1}{\beta} \, \sumint {\rm Tr} \left( \gamma^0
        S(q) \gamma^0 S(q) \right) \\
        &= -\frac{1}{\beta} \, \sumint {\rm Tr} \left( \gamma^0
        \frac{\partial}{\partial \mu} S \right) \\
        &= -\frac{\partial^2}{\partial \mu \partial \mu} \, \frac{1}{\beta} \, \sumint {\rm
        Tr} \, \ln S^{-1}.
    \end{split}
    \label{eq:kapusta2}
\end{equation}
Note that $S^{-1}(q) = (i \, \omega_n + \mu) \, \gamma^0 - \vec{q} \cdot \vec{\gamma} - M$, 
and we set $\mu \rightarrow 0$ after taking the derivatives. This yields an explicit
expression:

\begin{equation}
    \Pi_{00}^T(p^0=0, \vec{p} \rightarrow \vec{0}) = -\int \frac{d^3 q}{(2 \pi)^3} \, 4 \beta N_1 (1-N_1).
    \label{eq:electric_mass}
\end{equation}

\begin{figure}
	\resizebox{0.48\textwidth}{!}{%
	\includegraphics{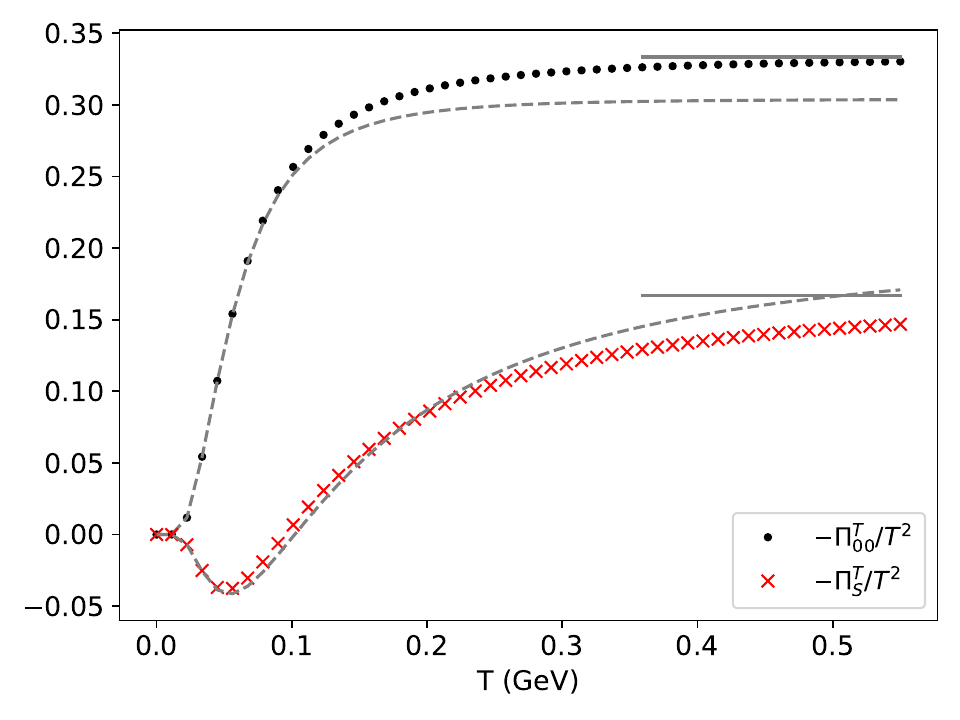}
	}
        \caption{The finite temperature polarization tensors
        (00-channel~\eqref{eq:electric_mass} and scalar~\eqref{eq:scalar_mass}) 
        in the static limit, normalized to $T^{-2}$ versus temperatures. We fix $M = 0.136 $
        GeV in this calculation.
        Dashed lines in gray are the corresponding quantities in the Boltzmann approximation. 
        Full lines denote massless limits. See text.
        }

\label{fig1}
\end{figure}

Similar analysis can be performed on other channels, e.g. for scalar and pseudoscalar cases:

\begin{equation}
    \label{eq:pis}
    \begin{split}
        \Pi_{S}(p^0, \vec{p}) &= -\frac{1}{\beta} \, \sumint {\rm Tr} \left(
        S(q) S(q+p) \right) \\
        &= 4 \left( I_0 + (\frac{1}{2} p^2-2 M^2) \, I_1 \right).
    \end{split}
\end{equation}

and

\begin{equation}
    \label{eq:pips}
    \begin{split}
        \Pi_{PS}(p^0, \vec{p}) &= \frac{1}{\beta} \, \sumint {\rm Tr} \left(
        \gamma_5
        S(q) \gamma_5 S(q+p) \right) \\
        &= 4 \left( I_0 + \frac{1}{2} p^2 \, I_1 \right).
    \end{split}
\end{equation}
The finite temperature contribution of scalar polarization can also be extracted 
by taking derivatives of pressure, now with respect to $M$ rather than to $\mu$:

\begin{equation}
    \label{eq:scalar_mass}
    \begin{split}
        \Pi_{S}^T(p^0=0, \vec{p} \rightarrow \vec{0}) &= 
        \frac{\partial^2}{\partial M \partial M} \, \frac{1}{\beta} \, 
        \sumint {\rm Tr} \, \ln S^{-1} \\
                &= -\int \frac{d^3 q}{(2 \pi)^3} \, 4 \beta \,(S_1-S_2) \\
            S_1 &= \frac{T}{E_1} (1-\frac{M^2}{E_1^2}) \, N_1 \\
            S_2 &= \frac{M^2}{E_1^2} \, N_1 (1-N_1).
    \end{split}
\end{equation}
which has the following limits: (1) at small $M$,

\begin{equation}
   \Pi_{S}^T(p^0=0, \vec{p} \rightarrow \vec{0}) \approx -\frac{1}{6} \, T^2
\end{equation}
verifying the low mass (or high temperature) expansion by Haber and
Weldon~\cite{weldon,Kapusta:2006pm};
and (2) the Boltzmann approximation,

\begin{equation}
    \begin{split}
        \Pi_{S}^T(p^0=0, \vec{p} \rightarrow \vec{0}) &\approx \frac{2}{\pi^2} \, M^2 \times \\
        &\quad \left( K_0(M/T) - \frac{T}{M}  K_1(M/T) \right),
    \end{split}
    \label{eq:msboltz}
\end{equation}
where $K_n$'s are the modified spherical Bessel function of the second kind.
Note the competition between the two terms in Eq.~\eqref{eq:msboltz}:
the latter, negative contribution dominates at $M/T \ll 1$, 
while the former positive contribution determines the $M/T \gg 1$ behavior.
This simply reflects the mass dependence of the thermal pressure of a free fermion gas $P_F$: 
while the pressure drops, at fixed $T$, when $M$ increases, 
the rate of change, reflected by $\frac{\partial^2}{\partial M \partial M} \, P_F$, 
starts being a negative value at small $M$, 
exhibits a peak at an intermediate $M$, 
and is suppressed (but with a positive value) at large $M$.

Lastly, we write down the corresponding results for $\Pi_{00}^T(p^0=0, \vec{p} \rightarrow
\vec{0})$: (1) at $M \rightarrow 0$ (or large $T$),
        \begin{equation}
\Pi_{00}^T(p^0=0, \vec{p} \rightarrow
\vec{0}) \approx -\frac{T^2}{3};
            \label{eq:lim1}
        \end{equation}
and (2) at large $M$ (or small $T$), where the Boltzmann approximation is valid,
        \begin{equation}
\Pi_{00}^T(p^0=0, \vec{p} \rightarrow
\vec{0}) \approx -\frac{2}{\pi^2} \, M^2 \,
            ( K_2(M/T)-K_2(2M/T) ).
            \label{eq:lim2}
        \end{equation}
In Fig.~\ref{fig1} we demonstrate a numerical calculation of these finite temperature
quantities. Various limits can be readily verified. 
Note that the scalar channel approaches the known high temperature limit 
substantially slower than the 00-channel. There, the Boltzmann approximation of the former
reaches $2/\pi^2 \times T^2$, 
compared to the full result of $1/6 \times T^2$.

\begin{figure*}
	\resizebox{\textwidth}{!}{
	\includegraphics{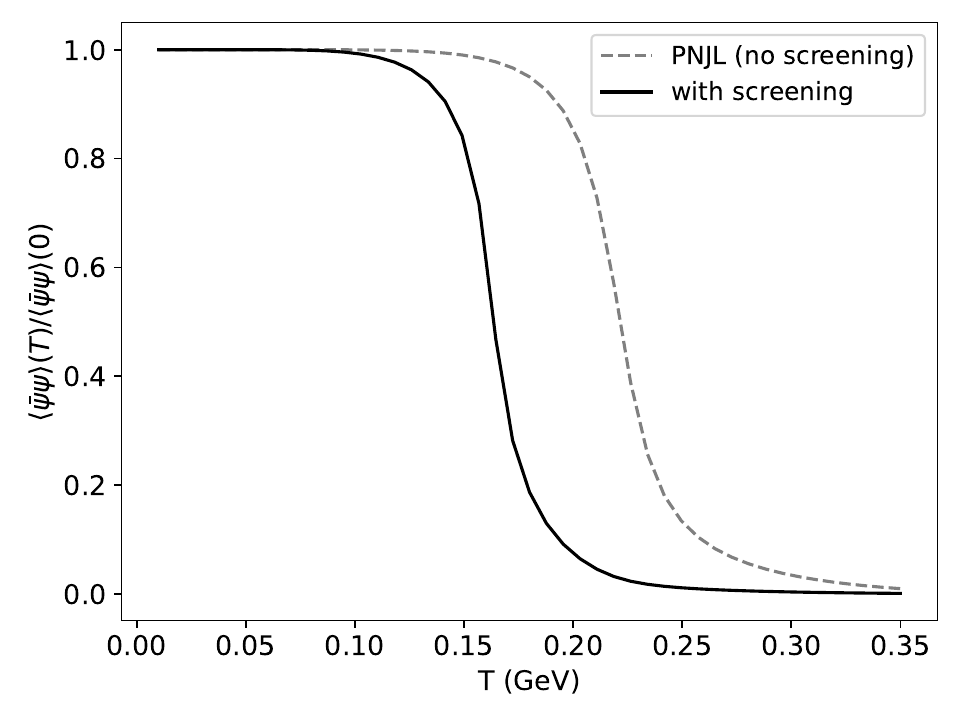}
	\includegraphics{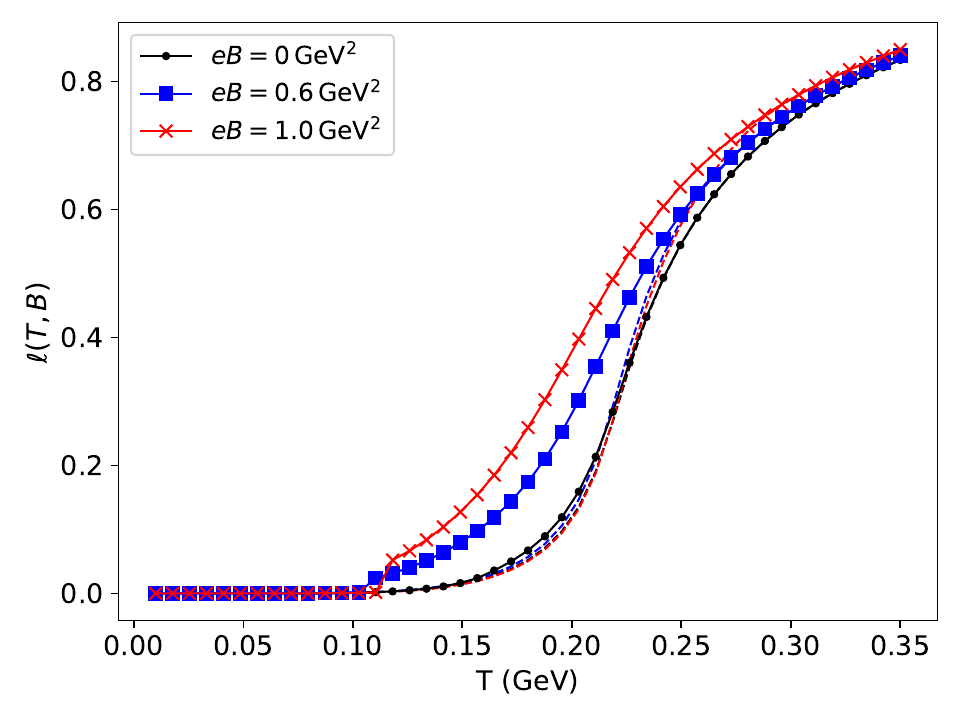}
	}
        \caption{The chiral condensate (normalized to the vacuum value) (left), and the
        Polyakov loop (right), versus the temperature, at finite magnetic field. 
        Dashed lines represent results obtained from a PNJL model with undressed coupling. 
        The model with dressed coupling is capable of producing the inverse magnetic catalysis at finite temperatures.}
\label{fig2}
\end{figure*}

\section{Results}

\subsection{Condensates and Polyakov loop}

Including the in-medium dressing by the $\Pi_{00}$ polarization tensor naturally relates the
deconfinement transition temperature and that of the chiral crossover transition. It also
plays a pivotal role in generating the inverse magnetic catalysis at finite temperatures
within the model. 

The results of chiral condensate have been presented in Ref.~\cite{new} and 
here we show the observables obtained under the Schwinger proper-time
regularization scheme. See Fig.~\ref{fig2} (left)
The results are similar to those obtained before in a 4D cutoff scheme~\cite{new}.

We highlight key theoretical features of the model:

    (1) A coupling to the Polyakov loop
    $\ell$~\cite{Fukushima:2003fw,Sasaki:2006ww,Fukushima:2017csk,Lo:2013hla,Lo:2014vba} is
    implemented in the model~\eqref{eq:cgap}.  This is done by
    replacing~\cite{Hansen} the thermal weight $N_{\rm th}(E)$ with ($N_c=3$)

\begin{equation}
    \begin{split}
N_{\rm th}(E) &\rightarrow \frac{1}{3} \, \sum_{j=1}^3
\frac{\hat{\ell}_F^{(j)}}{e^{\beta E}+\hat{\ell}_F^{(j)}} \\
        &= \frac{1}{3} \, \frac{3\ell \, e^{-\beta E} + 6 \ell \, e^{-2 \beta
        E} + 3 e^{-3\beta E}}{1 + 3 \ell \, e^{-\beta E} + 3 \ell \, e^{-2\beta E} + e^{-3\beta E}},
    \end{split}
    \label{eq:pls}
\end{equation}
where~\cite{Hansen,su3pot}
\begin{equation}
    \begin{split}
    \hat{\ell}_F &= {\rm diag} \, \left( e^{i \gamma_1}, 1, e^{-i
    \gamma_1} \right) \\
    \ell &= \frac{1}{3} \, {\rm Tr} \, \hat{\ell}_F = \frac{1}{3} \, (1 + 2 \cos \gamma_1).
    \end{split}
\end{equation}
The expectation value of the Polyakov loop needs to be determined from another
gap equation,

\begin{equation}
    \frac{\partial}{\partial \ell} \, (U_{\rm glue}(\ell) + U_Q(M, \ell)) = 0,
    \label{eq:newgap2}
\end{equation}
for a given pure gauge potential $U_{\rm glue}(\ell)$ and the quark potential
$U_{Q}(M, \ell)$, the latter describes the coupling of the Polyakov loop with quarks. 
These potentials have been studied extensively in
Refs~\cite{Fukushima:2003fw,Sasaki:2006ww,Lo:2013hla,Lo:2014vba,su3pot,Lo:2018wdo} and will not be repeated here.
In this work we employ the pure gauge potential in Ref.~\cite{Lo:2013hla}.

(2) The final set of gap equations for quarks becomes

\begin{equation}
        M = m + C_F \, \tilde{V}_0 \, M \times I_0(T; M, \ell),
    \label{eq:newgap1a}
\end{equation}
and

\begin{equation}
    \tilde{V}_0(T; M, \ell) = \frac{1} {V_0^{-1} - \frac{1}{2} \, N_f \, \Pi_{00}^T(T; M_0,
    \ell)}.
    \label{eq:newgap1b}
\end{equation}
As in Ref.~\cite{new}, we make a further approximation of using $M_0 = 0.136 $ GeV in the
ring. This point will be further improved in Sec.~\ref{sec4b}.
We have made explicit the dependence on temperature and the order parameter
fields $(M, \ell)$.

(3) The generalization of various quantities 
to a finite magnetic field $B$ can be implemented 
by replacing the momentum integral with a sum over the Landau levels~\cite{Andersen:2014xxa}:

\begin{equation}
    \int \frac{d^3 q}{(2 \pi)^3} \rightarrow \frac{\vert e_f \vert B}{2 \pi} \, \sum_{n=0}^\infty \, \frac{1}{2} \, \alpha_n \int_{-\infty}^{\infty} \frac{d q_z}{2 \pi}
\end{equation}
where $\alpha_n = 2 - \delta_{n0}$, 
$e_f$ is the electric charge of the species,
and replacing the transverse momentum by 

\begin{equation}
    q_x^2 + q_y^2 \rightarrow 2 \, n \times \vert e_f \vert  B.
\end{equation}
The modification of the integrals in Eq.~\eqref{eq:I_collect} is summarized 
in the appendix.

(4) One of the key objectives of this work is to examine the influence of ring
diagram on the Polyakov loop. (See Fig.~\ref{fig2} (right).)

The important observation is that the Polyakov loop becomes substantial at lower temperatures as magnetic field
increases, signaling lower transition temperature for deconfinement. 
This correct trend is brought forth by the polarization, and is quite robust against the use of different regularization schemes. 

To obtain known vacuum values of the physical observables: 
$f_\pi = 92.9$ MeV, $m_\pi = 137.8$ MeV, and
$\langle \bar{\psi} \psi \rangle = -(250 \, {\rm MeV})^3$ (per flavor) 
in the Schwinger proper-time regularization scheme, 
the set of model parameters is adjusted compared to Ref.~\cite{new}. 
They are given by $\Lambda = 1.101$ GeV, $G_{\rm NJL} \, \Lambda^2 = 3.668$ and a current
quark mass $m = 5$ MeV. We note that the constituent quark mass value in the Schwinger
scheme is substantially smaller ($\approx 200$ MeV) compared to the previous scheme
($\approx 300$ MeV) for the same value of chiral condensate.

(5) A positive feedback mechanism:
Solving Eqs.~\eqref{eq:newgap1a}, ~\eqref{eq:newgap1b} and ~\eqref{eq:newgap2} consistently, 
we obtain the results in Fig.~\ref{fig2}.
The reduction of the chiral transition temperature is obvious: 
the ring weakens the effective four-quark coupling at finite temperatures, 
leading to an earlier transition.
Also, without the polarization dressing in Eq.~\eqref{eq:dress},
the PNJL model predicts an increasing chiral transition temperature with $B$.
The dressed interaction reverses this trend, and the appearance of quarks enhances 
explicit Z(3) breaking, which weakens the confining effect and in turn enhances the
polarization~\cite{Lo:2020ptj}. This model demonstrates such a positive feedback mechanism in a very
transparent manner.

\begin{figure}
	\resizebox{0.5\textwidth}{!}{
	\includegraphics{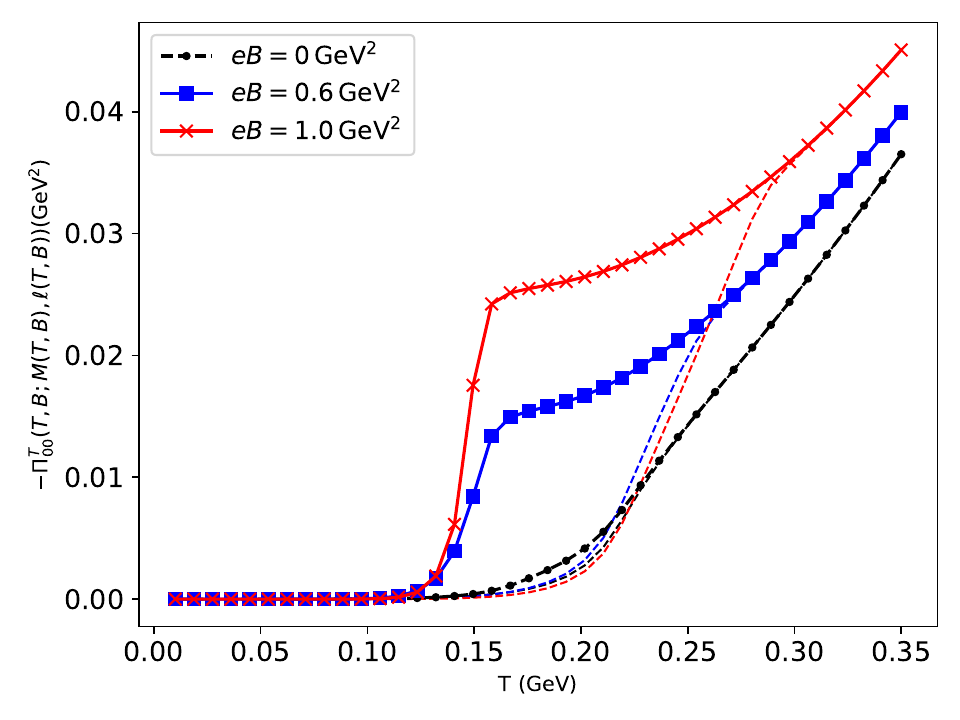}
	}
    \caption{The polarization tensor as a function of temperature, 
    evaluated with the quark mass and Polyakov loop determined from the gap equations.
    Dashed lines represent results obtained from a PNJL model with undressed coupling.}
\label{fig3}
\end{figure}

(6) Lastly we examine the effect of using a temperature (and $B$) dependent $(M, \ell)$, 
obtained as the solution to the gap equations, 
on the ring. This is shown in Fig.~\ref{fig3}.
The extra $(M, \ell)$ dependence turns out to give a modulation 
of the theoretical limits studied in Fig.~\ref{fig1}. 
Note that the polarization can increase substantially at intermediate temperatures 
with increasing $B$, which can drive the phenomenon of inverse magnetic catalysis.
We find no need to introduce a $B$-dependent coupling as advocated in Refs.~\cite{mediumG1,mediumG2}. 
Instead, Eq.~\eqref{eq:dress} could accommodate a theoretical explanation of such an effect.

\subsection{Truncation schemes}
\label{sec4b}

\begin{figure}
	\resizebox{0.5\textwidth}{!}{
	\includegraphics{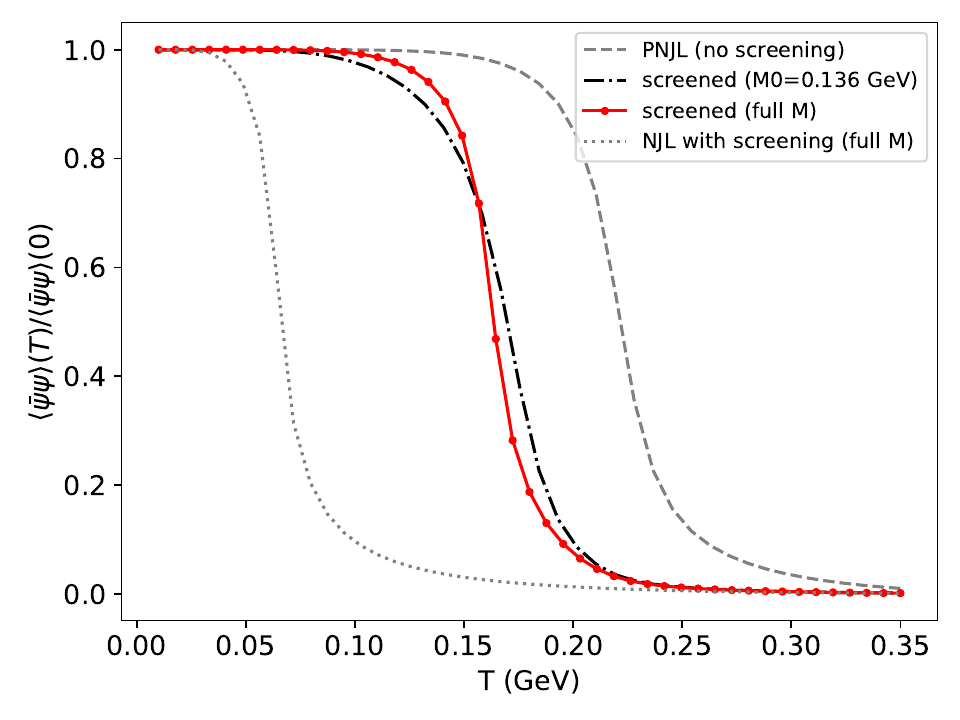}
	}
    \caption{The chiral condensate (with current quark mass contribution
        subtracted), normalized to the vacuum value, versus the temperature, at zero magnetic field. 
        Dashed (gray) line represents result obtained from a PNJL model without 
        screening effect. The dash-dotted (black) line is the result computed with the ring,
        under a further approximation of $M=M_0 = 0.136$ GeV. Such approximation is lifted
        in the fully consistent scheme ("full M").
        }
\label{fig4}
\end{figure}

It was reported in Ref.~\cite{new} that the use of full $M$ 
in $\tilde{V}_0(T; M, \ell)$ within a 4D cutoff scheme induces a first order phase transition, instead of the expected crossover
behavior. This is the reason why an extra condition $M = M_0$ is imposed. 
In this study, we find that the crossover nature of the transition is retained at $B=0$ when the Schwinger proper-time regularization scheme for the vacuum term is imposed.  
See Fig.~\ref{fig4}. 
Similar to the previous result, a transition temperature of
$\approx 160$ MeV is achieved, compared to $>200 $ MeV without the ring. 
This shows that the improvement is not constructed via a judicious choice of $M_0$, rather,
it is a natural result from iterating $M$ in the gap
equation~\eqref{eq:newgap1b}.~\footnote{In fact the results are not very sensitive to the
value of $M_0$ used.} 
The ability to link different scales naturally is one of the desirable feature of implementing the ring in
the chiral model. Finally, the dependence on cutoff scheme should motivate further study to
explore how the chiral phase transition depends on the assumed properties of gluons.

\subsection{Effect of Polyakov loop on ring}
\label{sec4c}

To illustrate the effect of the confinement, we perform the same
calculation while setting the value of the Polyakov loop field to unity, 
thus removing the confining effect on quarks. 
This leads to a dramatic decrease in the transition temperature, as shown in
Fig. ~\ref{fig4}. We have checked that such drastic change is
insensitive to the choice of the cutoff scheme.

Clearly, allowing the deconfined quarks in the ring diagram to dress the
4-point interaction at the low temperature phase gives a screening effect which
is too strong to produce an acceptable $T_c$. The coupling to the Polyakov loops, 
as shown in Eq.~\eqref{eq:pls}, remedies this problem. In addition, due to the
crossover nature of the transition, the chiral phase transition is now also
dependent on the details of the Polyakov loop potential. 

The problem of too strong screening by the quark loops has also been realized in
Ref.~\cite{Lo:2020ptj} even for a more elaborated model, giving transition
temperatures as low as $\sim 30$ MeV. The study here suggests that implementing the suppression of free, though massive, quarks with a
confining force in constructing the ring could provide a resolution.

\section{Conclusion}

In this work we have investigated the in-medium dressing of the four-quark interaction 
by the polarization. This provides a natural mechanism 
to resolve the problem of an overestimated chiral transition temperature in common PNJL models,
and is capable of generating an inverse magnetic catalysis at finite temperatures.
It is accomplished by a field theoretical incorporation of a quark loop
dressing and its feeding back to the quark gap equation. 
There is no need for artificial tuning of $T_d$ parameter in the gluon sector,
nor the need to introducing an explicit $B$-dependent coupling. Thus, the mechanism can
serve as a tentative explanation for the medium dependence discussed in the
literature.

Nevertheless, the current model makes some simplifying assumptions which
require improving. For example to make the problem more tractable we have employed the static approximation
of the ring. However dynamical (3-momentum dependences) and timelike (energy
dependence) effects can sometimes be drastic~\cite{Lo:2013lca}. 
Note that similar quark loops, in their timelike limits, are computed in
the model to derive pions and other mesons. 
In fact, it is an important question to understand how hadron loops enter in the
quarks-and-gluons-based picture. 
In principle, it is possible to understand, in accordance to quark-hadron duality, 
the former by including, and iterating, multi-particle interactions in the latter. 
Note that the role of gluon propagator, approximated as an
effective four-quark coupling, is formally recognized here. This is why quark
loop dressing~\eqref{eq:newgap1b} is introduced as an extension of standard mean-field results.
However, the current truncation scheme only includes these quark loops in dressing the coupling. 
It has yet to include additional interactions with the derived objects.

While we have explored the role of polarization in this work, vertex corrections are not
examined. The contact model is not ideal for this purpose, instead it
would be more satisfying to start with model which has a closer connection to
QCD. In addition, further work needs to be done to include an explicit treatment of dynamical gluons (and
ghosts) in the confinement model~\cite{Lo:2020ptj}. This gives a natural
extension to introduce non-local interactions among quarks, and allows to study the role
played by the gluons in a chiral phase transition. 
Finally we note an analogous dressing of the gluons is present at finite baryon density~\cite{eric_cgauge}. 
This could provide an additional handle to probe detailed features of the critical end
point~\cite{Ayala:2021nhx} predicted by the current model, and will be explored in the
future.

\begin{acknowledgement}
This study receives supports from the Polish National Science Center (NCN) under the Opus grant no. 2018/31/B/ST2/01663. M. S. acknowledges the support of the NCN Preludium grant no. 2020/37/N/ST2/00367.
\end{acknowledgement}

\section*{Appendix A: Finite $B$ integrals}
\label{appendix}

Here we collect the formulae of the integrals~\eqref{eq:I_collect} 
suitable for calculations at a finite magnetic field $B$.
For the following, we take $E_i = \sqrt{\vec{q}_i^2 + M^2}$ and $\vec{q}_1 = \vec{q}$ and $\vec{q}_2 = \vec{q}+\vec{p}$.
Starting with expression for $I_0$:

\begin{equation}
    \begin{split}
        I_0 &= I_0^{\rm vac} + I_0^{\rm vac, B} + I_0^{T, B} \\
        \\
        I_0^{\rm vac} &= \int \frac{d^3 q}{(2 \pi)^3} \, \frac{1}{2 E_1} \, \mathcal{R}_{\rm 3D}(q) \\
        \\
        I_0^{\rm vac, B} &= \lim_{\Lambda \rightarrow \infty} \left( S_0^{\rm vac, B} - I_0^{\rm vac} \right) \\
        S_0^{\rm vac, B} &= \frac{\vert e_f \vert B}{2 \pi} \, \sum_{n=0}^\infty \, \frac{1}{2} \, \alpha_n \, \int \, \frac{d q_z}{2 \pi} \, \frac{1}{2 E_1 } \, \mathcal{R}_{\rm 3D}(q) \\
        \\
        I_0^{T, B} &= \frac{\vert e_f \vert B}{2 \pi} \, \sum_{n=0}^\infty \, \frac{1}{2} \, \alpha_n \, \int \, \frac{d q_z}{2 \pi} \, \frac{-1}{2 E_1} \times 2 N_{\rm th}(E_1),
    \end{split}
\end{equation}
where
\begin{equation}
        E_1 = \sqrt{q_z^2 + 2 \,  n \,  \vert e_f \vert B + M^2}.
\end{equation}

The vacuum integral in the Schwinger proper-time regularization scheme reads
\begin{equation}
    I_0^{\rm vac} = \int_{1/\Lambda^2}^{\infty} \, \frac{dt}{(16 \pi^2)} \, \frac{1}{t^2} \, e^{-M^2 \, t}.
    \label{eq:schI0}
\end{equation}
An analytic expression for $ I_0^{\rm vac, B} $ can be derived from Refs.~\cite{Boomsma:2009yk,Lo:2020ptj}, 
it reads
\begin{equation}
    \begin{split}
    I_0^{\rm vac, B} &= \frac{M^2}{16 \pi^2} \times \\
     &\quad \left( \frac{\ln \Gamma(x_f)}{x_f} - \frac{\ln 2 \pi}{2 x_f} + 1 -
        (1-\frac{1}{2 x_f}) \ln x_f \right) \\
        x_f &= \frac{M^2}{ 2 \vert e_f \vert B}.
    \end{split}
    \label{eq:finiteB_I0}
\end{equation}

A similar analysis for $I_1$ gives:

\begin{equation}
    \begin{split}
        I_1 &= I_1^{\rm vac} + I_1^{\rm vac, B} + I_1^{T, B} \\
        \\
        I_1^{\rm vac}(p^0, \vec{p}) &= \int \frac{d^3 q}{(2 \pi)^3} \,
        \frac{-1}{4 E_1 E_2} \, \mathcal{R}_{\rm 3D}(q)) \times \\
        &\quad (\frac{1}{p^0-E_1-E_2}-\frac{1}{p^0+E_1+E_2} ) \\
        \\
        I_1^{\rm vac, B}(p^0, p_z) &= \lim_{\Lambda \rightarrow \infty} ( S_1^{\rm vac, B} - I_1^{\rm vac}) \\
        S_1^{\rm vac, B}(p^0, p_z) &= \frac{\vert e_f \vert B}{2 \pi} \, \sum_{n=0}^\infty \, \frac{1}{2} \, \alpha_n \, \int \, \frac{d q_z}{2 \pi} \, \frac{-1}{4 E_1 E_2} \, \mathcal{R}_{\rm 3D}(q) \, \\
            &\quad \times ( \frac{1}{p^0-E_1-E_2}-\frac{1}{p^0+E_1+E_2} ) \\
            \\
        I_1^{T, B} &= \frac{B}{2 \pi} \, \sum_{n=0}^\infty \, \frac{1}{2} \, \alpha_n \, \int \, \frac{d q_z}{2 \pi} \, 
        \frac{-1}{4 E_1 E_2} \times (Q^T_1 + Q^T_2), \\
        Q^T_1 &= -(N_1+N_2) \, \times \\
        &\quad (\frac{1}{p^0-E_1-E_2}-\frac{1}{p^0+E_1+E_2} ) \\
        Q^T_2 &= (N_1-N_2) \, \times \\
        &\quad (\frac{1}{p^0-E_1+E_2}-\frac{1}{p^0+E_1-E_2}).
    \end{split}
\end{equation}

We also record the vacuum result in the Schwinger proper-time regularization scheme:

\begin{equation}
    \label{eq:schI1}
    I_1^{\rm vac}(p)  = \int_0^1 \, dx \, \int_{1/\Lambda^2}^{\infty} \, \frac{dt}{(16 \pi^2 \, t)} \,
    \frac{1}{t} \, e^{-\tilde{M}^2 \, t},
\end{equation}
where
\begin{equation}
    \tilde{M}^2 - x(1-x) \, p^2.
\end{equation}

At vanishing external momentum, the vacuum integral $I_1^{\rm vac}(0, \vec{0})$ (the
timelike and spacelike limits coincide in vacuum) may be computed from a derivative relation:
\begin{equation}
    I_1^{\rm vac}(0, \vec{0}) = -\frac{1}{2 M} \, \frac{d}{d M} I_0^{\rm vac}.
    \label{eq:d-relation}
\end{equation}
Note how Eqs.~\eqref{eq:schI0} and~\eqref{eq:schI1} cleanly illustrate this relation.
Another application is to derive an analytic expression for $ I_1^{\rm vac, B}(0, \vec{0}) $
from Eq.~\eqref{eq:finiteB_I0}:
\begin{equation}
    \begin{split}
    I_1^{\rm vac, B}(0, \vec{0}) &= -\frac{1}{2 M} \, \frac{d}{d M} I_0^{\rm vac, B}
        \\
        &= \frac{1}{16 \pi^2} \times \left( -\psi(x_f+1) + \frac{1}{2 \, x_f} +
        \ln x_f \right),
    \end{split}
\end{equation}
where $\psi$ is the digamma function. This agrees with the result obtained in Ref.~\cite{Avancini:2015ady}. 

Finally we study the finite magnetic field extension of $\Pi_{00}^{T,B}$ in
Eq.~\eqref{eq:electric_mass} (per flavor):

\begin{equation}
    \begin{split}
        \Pi_{00}^{T,B}(p^0=0, \vec{p} \rightarrow \vec{0}) &= -\frac{\vert e_f \vert B}{2
        \pi} \, \sum_{n=0}^\infty \, \frac{1}{2} \, \alpha_n \\
        &\quad \times \int \, \frac{d q_z}{2 \pi} \, 4 \beta N_1 (1-N_1).
    \end{split}
    \label{eq:electric_massB}
\end{equation}
Examining in particular the contribution from the lowest Landau level (LLL) ($n=0$),
we get

\begin{equation}
    \begin{split}
    \Pi_{00}^{T,B}(p^0=0, \vec{p} \rightarrow \vec{0}) &= -\frac{\vert e_f \vert B}{4 \pi} \, \int
        \, \frac{d q_z}{2 \pi} \\
        &\quad \times \frac{4 \beta e^{\beta \sqrt{q_z^2+M^2}}}{(e^{\beta \sqrt{q_z^2+M^2}}+1)^2 },
    \end{split}
\end{equation}
which for massless quarks reduces to

\begin{equation}
\Pi_{00}^{T,B}(p^0=0, \vec{p} \rightarrow \vec{0}) = -\frac{\vert e_f \vert B}{2 \pi^2}.
    \label{eq:LLL}
\end{equation}
This gives an alternative derivation of the result in Ref.~\cite{debye,Alexandre:2000jc}.
A similar integral appears in the study of the explicit Z(3) symmetry breaking~\cite{Lo:2020ptj}.

\bibliographystyle{unsrt}
\bibliography{ref}

\end{document}